\documentclass[aps,prd,reprint,nofootinbib,superscriptaddress,preprintnumbers,amsmath,amssymb,latexsym,array,enumerate,letterpaper,twocolumn]{revtex4-1}

\usepackage{inputenc,amsmath,amssymb}
\usepackage{graphicx}
\usepackage{natbib}
\usepackage{xcolor}
\usepackage[colorlinks=true,urlcolor=brown,linkcolor=blue,citecolor=magenta]{hyperref}
\usepackage{graphicx}
\usepackage{comment}
\usepackage{booktabs}

\usepackage{soul}

\newcommand{\beq}{\begin{eqnarray}}
\newcommand{\eeq}{\end{eqnarray}}
\newcommand{\bea}{\begin{eqnarray}}
\newcommand{\eea}{\end{eqnarray}}
\newcommand{\bwt}{\begin{widetext}}
\newcommand{\ewt}{\end{widetext}}

\newcommand{\eV}{\textrm{ eV}}

\newcommand{\GeV}{\textrm{ GeV}}

\begin{document}

\title{Spectral distortions of astrophysical blackbodies as axion probes}

\author{Jae Hyeok Chang}
\email{jaechang@umd.edu}
\affiliation{Department of Physics and Astronomy, Johns Hopkins University, Baltimore, MD 21218, USA}
\affiliation{Maryland Center for Fundamental Physics, Department of Physics, University of Maryland, College Park, MD 20742, USA}

\author{Reza Ebadi}
\email{ebadi@umd.edu}
\affiliation{Department of Physics, University of Maryland, College Park, Maryland 20742, USA}
\affiliation{Quantum Technology Center, University of Maryland, College Park, Maryland 20742, USA}

\author{Xuheng Luo}
\email{xluo26@jh.edu}
\affiliation{Department of Physics and Astronomy, Johns Hopkins University, Baltimore, MD 21218, USA}

\author{Erwin H. Tanin}
\email{ehtanin@gmail.com}
\affiliation{Department of Physics and Astronomy, Johns Hopkins University, Baltimore, MD 21218, USA}

\preprint{UMD-PP-022-05}

\begin{abstract}
     Recent studies reveal that more than a dozen of white dwarfs displaying near-perfect blackbody spectra in the optical range have been lurking in the Sloan Digital Sky Survey catalog. We point out that, in a way analogous to the Cosmic Microwave Background, these stars serve as excellent testbeds for new physics. Specifically, we show how their observed lack of spectral distortions translates into limits on the parameter space of axions with electromagnetic coupling. The prospects for future improvements are also discussed. 
\end{abstract}

\maketitle

\section{Introduction}
All blackbodies at a given temperature emit the same radiation spectrum, regardless of their material composition and pre-equilibrium history. This universal and well-understood feature of ideal blackbodies makes them excellent probes of beyond the Standard Model (BSM) physics which generically causes spectral distortions. A prime example is the Cosmic Microwave Background (CMB), whose observed near-perfect blackbody spectrum has been used to place stringent limits on axions \cite{Mirizzi:2009nq,Cadamuro:2011fd,Mukherjee:2018oeb,Bolliet:2020ofj,Balazs:2022tjl}, dark photons \cite{Jaeckel:2008fi,Mirizzi:2009iz, McDermott:2019lch, Caputo:2020bdy,Caputo:2020rnx}, millicharged particles \cite{Melchiorri:2007sq, Berlin:2022hmt}, and sterile neutrinos \cite{Mirizzi:2007jd}; see \cite{Chluba:2013vsa, Chluba:2015hma, Ali-Haimoud:2015pwa,Poulin:2016anj, Acharya:2018iwh, Liu:2023fgu,Liu:2023nct} for model-independent studies and \cite{Lucca:2019rxf, Chluba:2019kpb,Chluba:2019nxa} for reviews. In this work, we propose an astrophysical analog of this technique.

The radiation spectrum of a generic star roughly follows a blackbody profile, however non-trivial deviations are expected for the following reasons. While photons trapped in the interior of a star are locally thermalized and therefore follow a blackbody spectrum, the photons that we observe are those that escape from the interface of the star where the local environment is departing from thermal equilibrium. These decoupled photons would then go through the outer semi-transparent layers of the stellar atmosphere and have their spectrum shaped by opacity effects, which often include various atomic lines and molecular bands. Hence, the spectrum emerging from a star \textit{usually} does not resemble a blackbody. Having said that, the state of local thermal equilibrium with its associated blackbody radiation spectrum is an attractor for many interacting systems. If a star with a high-quality blackbody spectrum is observed, there is likely an underlying physical reason.

Recent studies \cite{2018AJ....156..219S, 2019A&A...623A.177S, 2020IAUGA..30..480F} found 17 stars in the Sloan Digital Sky Survey (SDSS) catalog, dubbed \textit{blackbody stars}, that display almost perfect blackbody spectra with no apparent spectral lines in the optical regime. The brightness, spectra, and Gaia parallaxes of these stars imply that they are DC-type white dwarfs with effective temperatures and distances lying in the range $6600-11000\text{ K}$ and $71-220\text{ pc}$, respectively. Since Balmer lines would be visible at such temperatures if the atmospheres of these stars were hydrogen-rich, the absence of lines suggest that these stars are helium-rich \cite{2019ApJ...878...63B, Saumon:2022gtu}.\footnote{The strong gravity on the surface of a white dwarf implies that only the lightest elements, namely hydrogen and helium, would stay afloat.} Incidentally, the opacity of a helium-rich atmosphere with aforementioned temperatures is nearly frequency-independent in the optical range, which explains the observed blackbody-like spectrum \cite{2006ApJ...651L.137K,2009AJ....138..102K, 2009ApJ...696.2094K}.\footnote{The atmospheres of the blackbody stars are currently best explained as being dominated by helium but with a tiny hydrogen contamination \cite{2010ApJS..190...77K, 2018AJ....156..219S, 2019A&A...623A.177S}. While pure-helium model atmospheres provide a good fit to the blackbody-like spectra, these models also imply a surface gravity that is unrealistically strong for a white dwarf. The lack of detectable spectral lines in the spectra of blackbody stars act as a constraint on but do not preclude the presence of hydrogen contamination with spectral lines weak enough to be hidden in the uncertainties. Such a tiny amount of hydrogen can in fact change the continuum spectrum slightly, leading to a lower and more realistic surface gravity. The inclusion of hydrogen improves the fits to the spectra and the a priori unknown H/He ratio are determined based on the goodness of fit to be $\sim 10^{-6}$ \cite{2018AJ....156..219S}.} Aside from opacity effects, atmospheric stratification may slightly broaden the spectrum relative to that of a pure blackbody. We discuss this point in Appendix~\ref{appendix:gray}.

Assuming our Standard Model understanding of the blackbody-star spectra is correct, we can probe BSM physics that affects these spectra. For definiteness, we adopt the highly popular model of the axionlike particle, namely a scalar $a$ with two-photon coupling $(g_{a\gamma\gamma}/4)F\tilde{F}$, as a convenient benchmark to compare our technique against other proposals. Throughout the paper, we will refer to this scalar simply as the axion. The strong magnetic fields present in a wide range of astrophysical environments, including those on the paths to the blackbody stars, enable photons to convert into axions \cite{Raffelt:1987im, Raffelt:1996wa, Marsh:2021ajy} and thereby distort the spectra of the blackbody stars. We can probe an axion parameter space based on how well the spectrum predicted by that parameter space fit the observed data.

Works in the same spirit as ours have previously appeared in the literature \cite{Chelouche:2008ta, AxionLimits}. Assorted astrophysical objects including isolated neutron stars \cite{Zhuravlev:2021fvm,Zhuravlev:2021mum}, active galactic nuclei \cite{Fermi-LAT:2016nkz,Berg:2016ese,Chen:2017mjf, Libanov:2019fzq,Reynolds:2019uqt}, blazars and quasars \cite{Fairbairn:2009zi,Galanti:2018upl}, and SN1987A \cite{Payez:2014xsa} have been used as backlights for spectral distortion effects; see also \cite{Grin:2006aw,Regis:2020fhw,Bessho:2022yyu} which look for spectral signals from a relic population of axions. These studies rely on sufficient understanding of the relevant astrophysical environments and thus suffer from various modelling uncertainties. While systematics are inevitably present in our work as well, the strength of our method lies in the quality of the blackbody spectra under consideration and the prospects for future improvements.

In what follows, we first characterize and parameterize the spectral-distortion signals due to photon-axion conversions in Section~\ref{s:signals}, both in the local galactic magnetic field (Section~\ref{ss: galacticconversion}) and in putative blackbody-star magnetic fields (Section~\ref{ss:stellarconversion}). We then place limits on the axion parameter space based on the overall goodness of fit of the model-predicted spectra to the observed blackbody-star spectra as well as discuss the future prospects of the proposed technique in Section~\ref{s:results}, before concluding in Section~\ref{s:conclusion}.

\section{Axion spectral-distortion signals}
\label{s:signals}

\begin{figure}
    \centering
    \includegraphics[width=\linewidth]{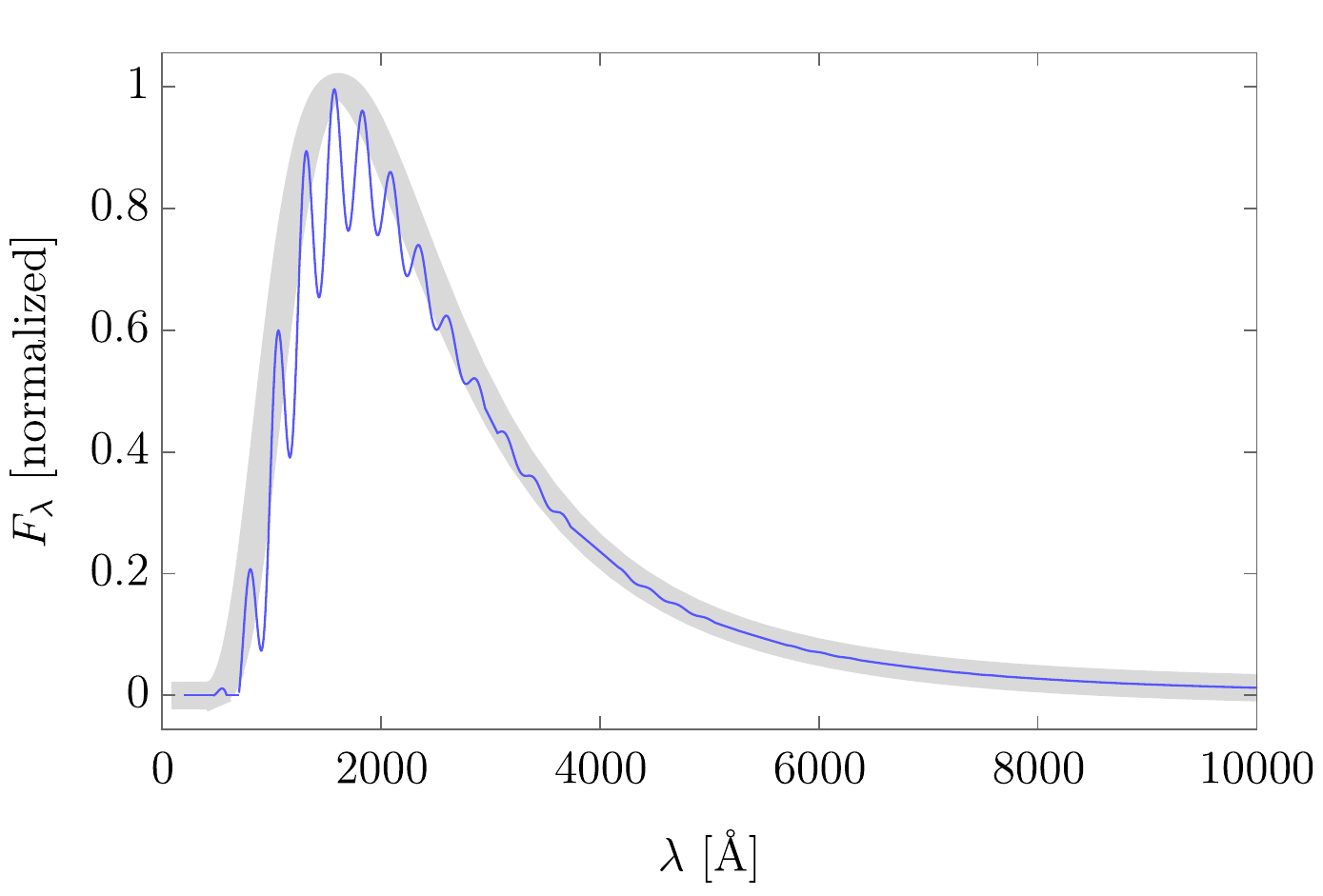}
    \caption{Schematic visualization of spectral-distortion effects due to photons converting into axions. The gray band is a sketch of the observed near-perfect blackbody flux-per-unit-wavelength $F_\lambda$ data points, with its thickness representing the error bars. The blue line depicts the predicted $F_\lambda$ shape in the presence of axion.}
    \label{fig:BBdistortion}
\end{figure}

To demonstrate the feasibility of our proposed approach, we consider as an example the photon disappearance signals predicted by the existence of an axion $a$, i.e. a pseudoscalar defined by the following Lagrangian:
\begin{align}
    \mathcal{L}_a=\frac{1}{2}(\partial a)^2-\frac{1}{2}m_a^2 a^2-\frac{g_{a\gamma\gamma}}{4}a F\tilde{F}
\end{align}
with the mass $m_a$ and photon coupling $g_{a\gamma\gamma}$ completely unrelated. In the presence of an external magnetic field $\vec{B}_{\rm ext}$, the axion-photon coupling term $-(g_{a\gamma\gamma}/4)aF\tilde{F}$ 
turns (in the Coulomb gauge $\vec{\nabla}.\vec{A}=0$) into an effective mass-mixing term between the axion $a$ and the transverse (to the wavevector) photon polarization $A_\parallel$ that is parallel to $\vec{B}_{\rm ext}$, namely $\sim (g_{a\gamma\gamma}B_{\rm ext}/\lambda) a A_{\parallel}$, which is wavelength $\lambda$ dependent. As a result, photons propagating in a magnetic field background over some distance may convert to axions with wavelength-dependent probabilities $P_{\gamma\rightarrow a}(\lambda)$, which may in turn lead to detectable distortions in the continuum spectra of the blackbody stars (see Fig.~\ref{fig:BBdistortion}). We can claim a discovery of the axion if spectral distortions with frequency-dependence matching that from photon-axion conversion are detected at high significance or, conversely, place limits on the axion parameter space in the absence of irregularities in the spectra.

In cases of our interest, the photon to axion conversion probability in a medium with spatially-varying plasma frequency $\omega_p$ and magnetic field orthogonal to the photon trajectory $B$ can be computed with the following formula \cite{Dessert:2019sgw}
\begin{align}
    P_{\gamma\rightarrow a}=\frac{1}{2}\left|\frac{g_{a\gamma\gamma}}{2}\int_0^d dz' B(z')e^{i\int_0^z dz'' \frac{\omega_p^2(z'')-m_a^2}{4\pi}\lambda}\right|^2 \label{weakmixing}
\end{align}
where $z$ and $d$ denote, respectively, arbitrary and total distance traversed by the photon. Note that the extra factor of $1/2$ corresponds to the case where only one of the two linear polarization states of a photon mixes with the axion. The above formula assumes relativistic photons/axions and weak photon-axion mixing $P_{\gamma\rightarrow a}\ll 1$. The former is easily satisfied since we are mainly concerned with optical energy ranges (around the$\eV$ scale) which is far above the relevant $\omega_p$ in all the cases we consider, while the latter is justified since the axion can only be responsible for small spectral distortions in a spectrum that is already observed to be a near-perfect blackbody.

The photons emitted by a blackbody star can convert to axions in the vicinity of the star facilitated by the stellar magnetic field \cite{Dessert:2019sgw, Dessert:2021bkv, Dessert:2022yqq} and/or in the interstellar medium facilitated by the Galactic magnetic field \cite{Mukherjee:2018oeb, Xiao:2020pra}. In what follows, we compute the predicted conversion probabilities in both cases using the approximate conversion probability formula \eqref{weakmixing}. The results are shown in Figs.~\ref{fig:galconvprob} and \ref{fig:WDconvprob} for several fiducial cases. Note that the rich features predicted by the axion physics are distinct from those expected from the opacity of a white dwarf atmosphere, providing a powerful handle to detect or constrain axions.

\subsection{Photon-axion conversion in the galactic magnetic field}
\label{ss: galacticconversion}

\begin{figure*}
    \centering
    \includegraphics[width=0.48\linewidth]{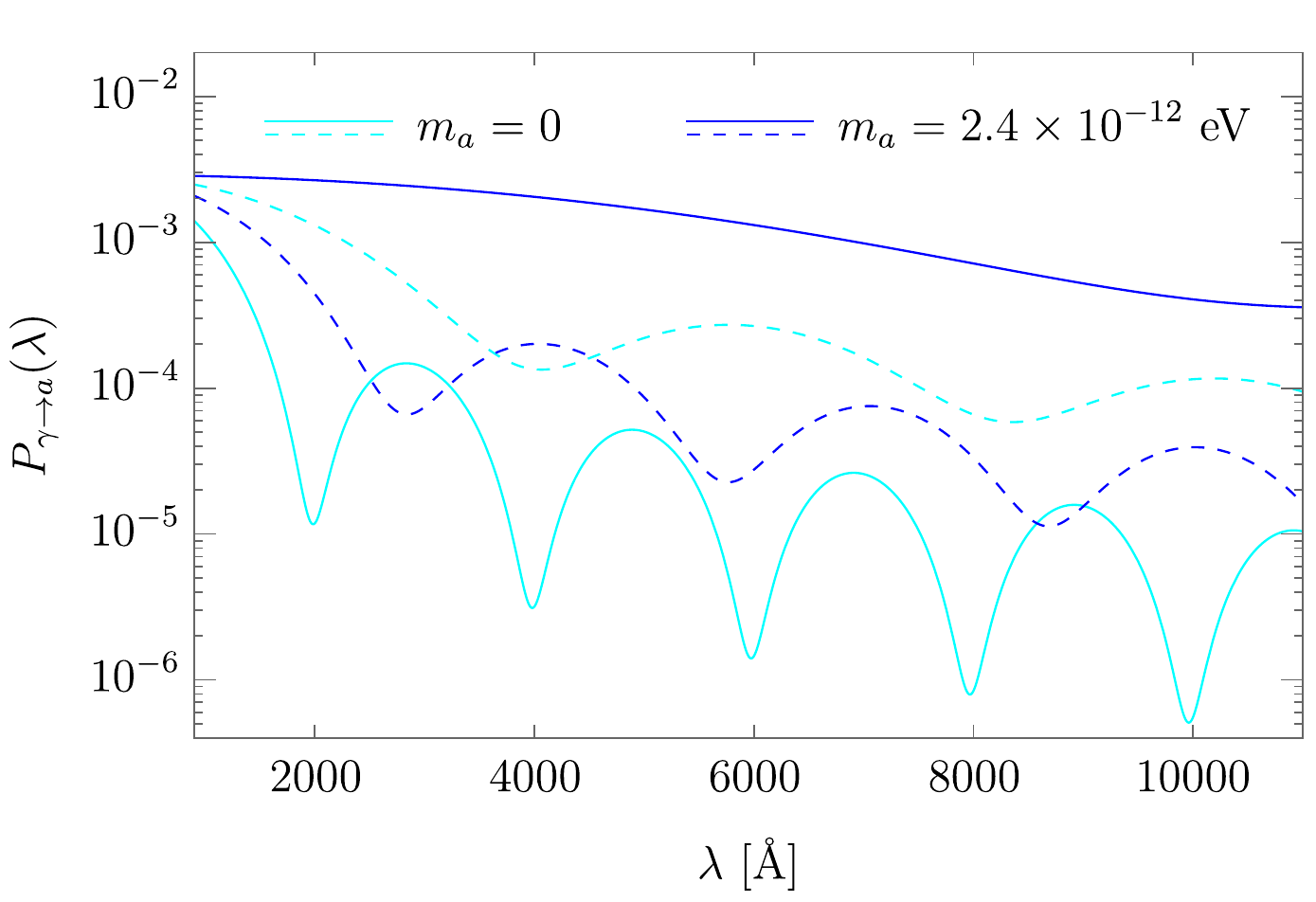}~~~
    \includegraphics[width=0.47\linewidth]{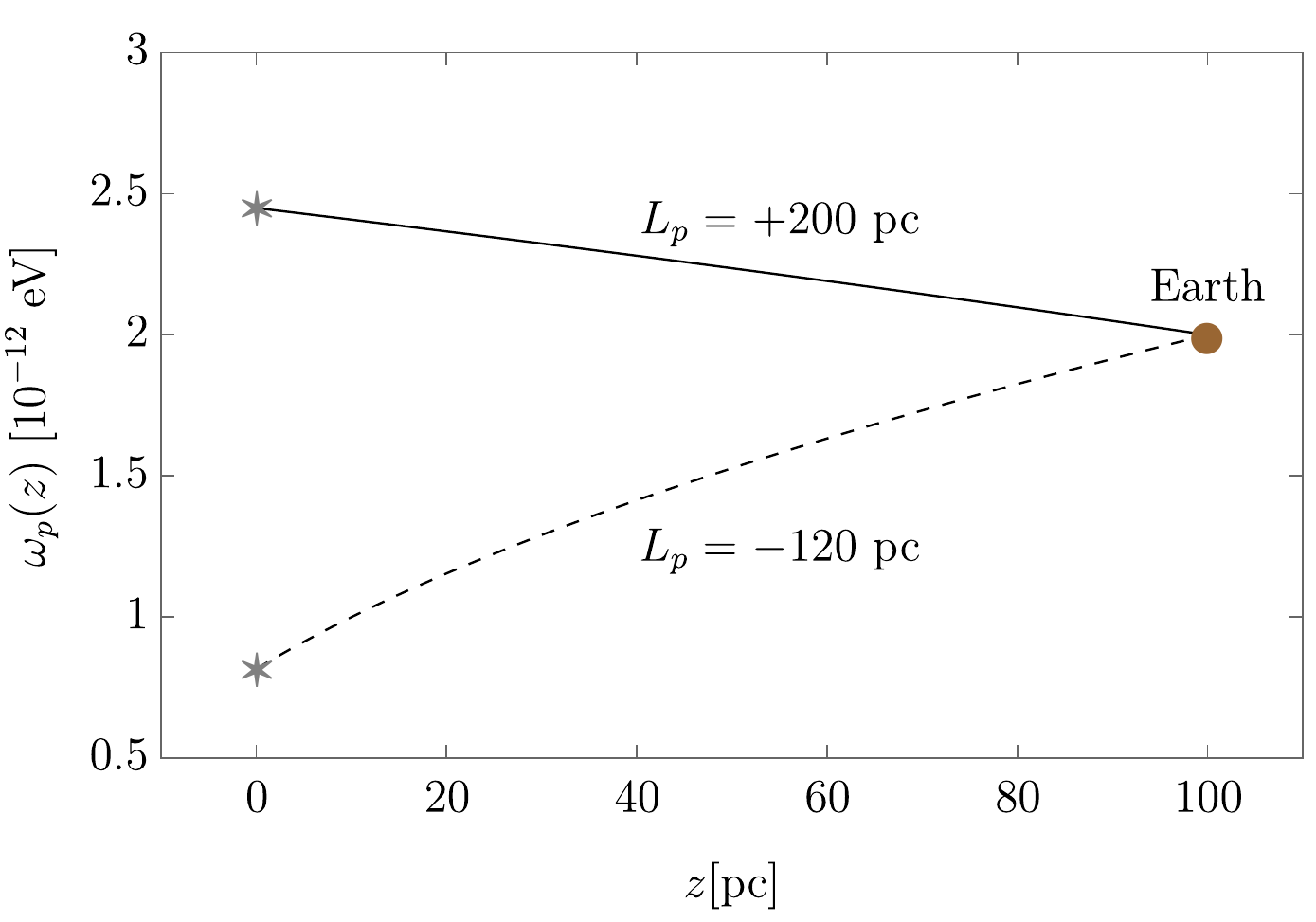}
    \caption{\textit{Left: }the probability as a function of wavelength of photons emitted by a star to convert into axions enroute to the Earth $P_{\gamma\rightarrow a}(\lambda)$, c.f.~\eqref{Pgal}, via the (assumed-uniform) galactic magnetic field $B_{\rm gal}=5\,\mu\text{G}$. The solid and dashed lines refer to different assumptions on the line-of-sight plasma mass profile as depicted in the right figure. \textit{Right: }the assumed plasma mass profiles over the distance $z$ traversed by the photons $\omega_p(z)$, given by \eqref{plasmamassprofile} for $L_p=+200\text{ pc}$ (solid) and $L_p=-120\text{ pc}$ (dashed).
    Different $L_p$'s represent generically different plasma mass profiles experienced by photons emitted from different stars.
    Here, the axion-photon coupling is $g_{a\gamma\gamma}=10^{-10}\GeV^{-1}$ and the distances to both stars are set to $d=100\text{ pc}$. Of the four cases shown, only the case with $m_a=2.4\times 10^{-12}\eV$ and $L_p=+200\text{ pc}$ has an occurence of resonance crossing, $\omega_p(z)=m_a$, which explains why the $P_{\gamma\rightarrow a}$ in that case is the highest.}
    \label{fig:galconvprob}
\end{figure*}

The solar system is inside a region of the interstellar medium called the Local Bubble, characterized by its relatively low density, high temperature, and high ionization. Depending on which direction we look, the inner boundary of the Local Bubble lies $\sim 50-200\text{ pc}$ from where we are, meaning that most but not all of the blackbody stars are inside the Local Bubble. The magnetic field and plasma mass profiles in the Local Bubble do not follow that of the typical large-scale galactic magnetic field. A simplified model for the electron density profile in the Local Bubble \cite{2017ApJ...835...29Y} suggests that the electron density $n_e$, and hence the plasma mass $\omega_p$, varies by $O(1)$ over a length scale of $10-100\text{ pc}$ depending on the direction, with a local mean value of $n_e\approx 5\times 10^{-3}\text{ cm}^{-3}$, which corresponds to $\omega_p\approx 2\times 10^{-12}\eV$.

For the purpose of computing the photon-axion conversion probability, we model the Local Bubble as a region with a uniform static magnetic field, with the component relevant for photon-axion mixing  $B_{\rm gal}$ taken to be \cite{Krause:2021xav}
    \begin{align}
        B_{\rm gal}=5\,\mu\text{G} \label{Bgal}
    \end{align}
and a linear plasma mass squared profile along the photon/axion trajectory
\begin{align}
    \omega_p^2(z)=\omega_{p,0}^2\left(1+\frac{d-z}{L_{p}}\right) \label{plasmamassprofile}
\end{align}
where $\omega_{p,0}=2\times 10^{-12}\eV$ is the plasma mass in the vicinity of the Earth \cite{Ocker:2020tnt}, $\omega_{p,0}^2/L_p$ is the assumed-constant local gradient of $\omega_{p}^2$ along a line of sight, and $z$ is the photon/axion displacement relative to a star and toward the Earth. This approximation is reasonable if the actual $\omega_p$ profile is approximately monotonic in a direction of interest. We expect $L_{p}$ to be of order $10-100\text{ pc}$ in magnitude and can be positive or negative (keeping $\omega_p^2(z)>0$) \cite{2017ApJ...835...29Y}. This simple model of the Local Bubble allows us to analytically capture both non-resonant conversions (which depend on the average $B_{\rm gal}$ and $\omega_p$) as well as resonant conversions (which depend on not only the average $B_{\rm gal}$ and $\omega_p$, but also the gradient $\partial_z\omega_p$ at resonance, here parameterized by $L_p$).

Assuming the local galactic magnetic field and plasma mass profile are given by \eqref{Bgal} and \eqref{plasmamassprofile}, the conversion probability can be computed exactly \cite{Marsh:2021ajy}
\begin{align}
P_{\gamma\rightarrow a}&=\frac{1}{2}\frac{\pi^2 g_{a\gamma\gamma}^2B_{\rm gal}^2L_p}{2\omega_{p,0}^2\lambda}\Bigg|
\text{Erf}\left[\Phi(d)\right]-\text{Erf}\left[\Phi(0)\right]\Bigg|^2\\
\Phi(z)&=\sqrt{\frac{iL_p \lambda}{2\pi}}\frac{m_a^2-\omega_p^2(z)}{2\omega_{p,0}} \label{Pgal}
\end{align}
As shown in Fig.~\ref{fig:galconvprob}, this conversion probability displays several notable spectral features which depend on the parameter space. One can spot a relatively fast oscillatory pattern over wavelength  with periodicity $\Delta \lambda$ around the slowly-varying, local-average value of the probability $\bar{P}_{\gamma\rightarrow a}(\lambda)$. The periodicity $\Delta \lambda$ is set by the typical value of $\omega_{p,0}^2/[(m_a^2-\omega_p^2(z))^2L_p]$, while the slowly-varying probability is roughly $\bar{P}_{\gamma\rightarrow a}(\lambda)\sim [g_{a\gamma\gamma}B_{\rm gal}(L_p/\omega_{p,0}^2\lambda)]^2$ if $\omega_p(z)$ crosses $m_a$ at some point (i.e. a resonance occurs) and is highly suppressed otherwise. Hence, the strongest conversion probabilities are expected near $m_a\approx \omega_{p,0}=2\times 10^{-12}\eV$.

\subsection{Photon-axion conversion in the stellar magnetic field}
\label{ss:stellarconversion}

\begin{figure}
    \centering
    \includegraphics[width=\linewidth]{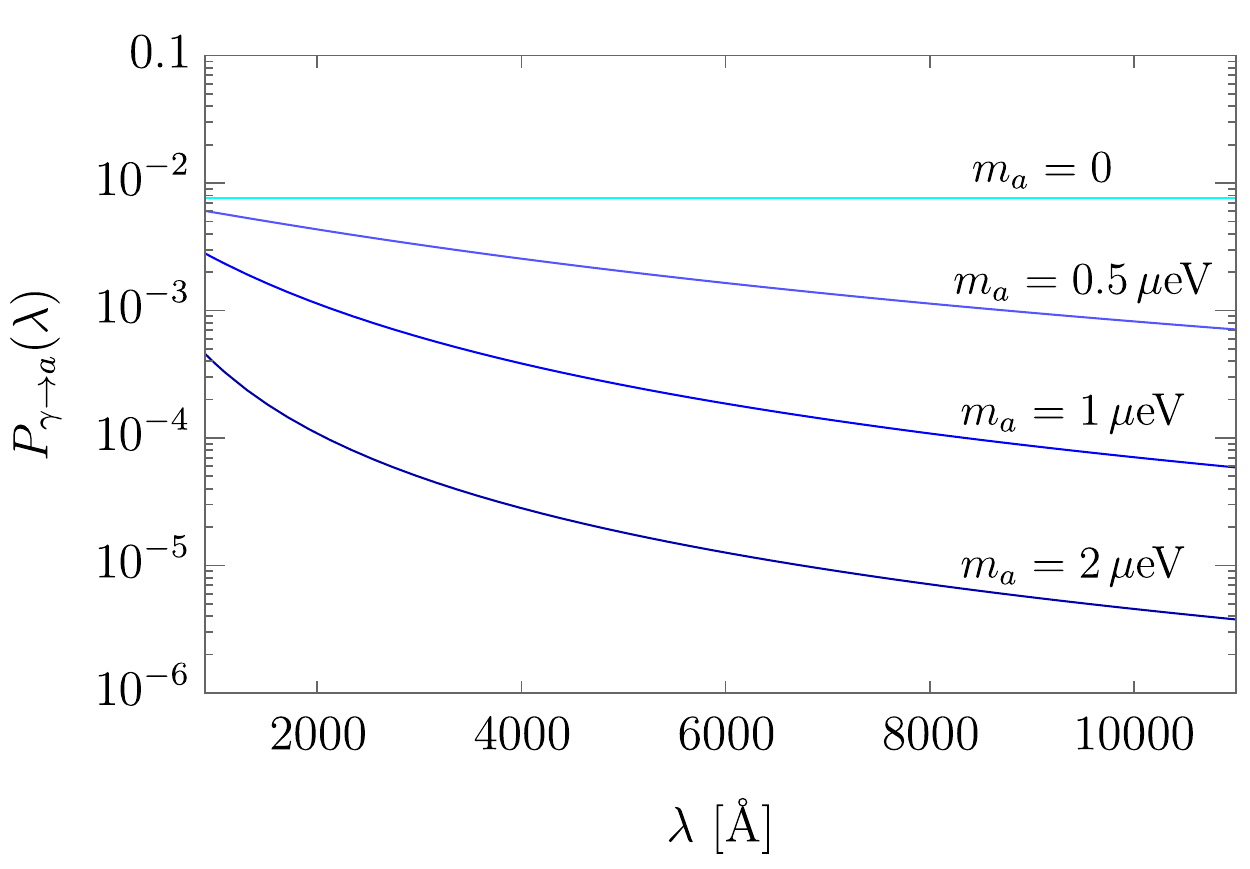}\\
    \caption{Photon to axion conversion probability as a function of wavelength $P_{\gamma\rightarrow a}(\lambda)$, c.f. \eqref{PWD}, via the blackbody star (white dwarf) magnetic field for different axion masses $m_a$. Here, the axion-photon coupling, stellar magnetic field at the magnetic pole, and radius are set to $g_{a\gamma\gamma}=10^{-11}\text{ GeV}^{-1}$, $B_{\rm WD}=10^{8}\text{ G}$, and $R_{\rm WD}=10^4\text{ km}$, respectively. We assume that the viewing angle relative to the pole is $\theta=\pi/2$ which corresponds to $F(\pi/2)=1$ in \eqref{PWD}.}
    \label{fig:WDconvprob}
\end{figure}

The magnetic fields of the blackbody stars have not been measured. However, as we will argue below, statistically speaking we expect several of them to have a strong magnetic field in the range $\sim 10^{4}-10^{9}\text{ G}$. Pending actual magnetic field measurements, we derive the axion conversion signals for a putative magnetic field in the aforementioned range.

The magnetic field at a position $\vec{r}=r\hat{r}$ from the center of a white dwarf can be modeled as that due to a dipole pointing at $\hat{m}$
\begin{align}
    \vec{B}(\vec{r})= \frac{B_{\rm WD} R_{\rm WD}^3}{2r^3}(3(\hat{m}\cdot \hat{r})\hat{r}-\hat{m})
\end{align}
where $R_{\rm WD}$ and $B_{\rm WD}$ are the radius and north pole magnetic field of the white dwarf. Neglecting the plasma frequency and strong QED effects,\footnote{The plasma frequency is negligible as it leads to negligible phase change compared to the axion mass in the conversion amplitude \eqref{weakmixing} for axion masses $m_a\sim \mu eV$, where the axion-induced spectral distortions are the strongest. Strong QED effects \cite{Chelouche:2008ta, Dessert:2022yqq}, i.e. corrections to the photon dispersion relation due to virtual electron-positron pairs, may also lead to an additional phase in the conversion amplitude in the presence of strong background magnetic fields. 
We have checked that the latter is negligible as long as the white dwarf magnetic field does not exceed $B_{\rm WD}\sim 10^9\text{ G}$.} it can be shown that the $P_{\gamma\rightarrow a}$ in the vicinity of a white dwarf reduces to 
\begin{align}
    P_{\gamma\rightarrow a}\approx & \frac{1}{2}F(\theta)\frac{(g_{a\gamma\gamma}B_{\rm WD}R_{\rm WD})^2}{16}\left|\int_{1}^{\infty}d\tilde{r}\frac{e^{i\delta_a\tilde{r}}}{\tilde{r}^3}\right|^2 \label{PWD}
\end{align}
where $\delta_a=-m_a^2R_{\rm WD}\lambda/4\pi$, $\cos\theta=\hat{m}.\hat{n}_{\rm obs}$ with $\hat{n}_{\rm obs}$ being the line-of-sight direction, and $F(\theta)$ is a numerical factor in the approximate range $F(\theta)\approx 1-2$ which quantifies the $\hat{n}_{\rm obs}$ dependence of the probability. As shown in Fig.~\ref{fig:WDconvprob}, the wavelength dependence of the conversion probability is generally monotonic. The conversion probability is maximized when $|\delta_a|$ is minimized (in which case the integral in \eqref{PWD} adds up in phase), however in this case the probability is essentially flat over wavelengths, resulting in a photon disappearance signal that is degenerate with the radius and distance of the white dwarf. We expect the most apparent distortion signal (and hence the best axion limits) to occur for a typical white dwarf at axion masses $m_a\sim\mu\text{eV}$ corresponding to $|\delta_a|\sim 1$
, which result in conversion probabilities that are not only unsuppressed but also have significant variation over the wavelength window of interest.

For radially moving photons in the direction $\hat{r}=\hat{n}_{\rm obs}$, one can calculate $P_{\gamma\rightarrow a}$ exactly, yielding \eqref{PWD} with $F(\theta)=\sin^2\theta$. However, a photon emitted from a generic point on the white dwarf surface $R_{\rm WD}\hat{r}_{\rm emit}$ moves in the direction $\hat{n}_{\rm obs}\neq \hat{r}_{\rm emit}$ in order to reach our telescope, i.e. the majority of the photons do not move radially. To obtain the appropriate conversion probability for our purposes, one must perform averaging over all the possible trajectories from the white dwarf hemisphere facing our telescope as well as both transverse photon polarizations. We have done so numerically (see Appendix~\ref{appendix:nonradial}) and found that while there are deviations, the dependence of the averaged $P_{\gamma\rightarrow a}$ on parameters other than the angle $\theta$ closely resembles the radial one, namely \eqref{PWD}. However, unlike in the radial case, the $F(\theta)$ now varies monotonically with $\theta$ from $F(\theta=0)\approx 2$ where it is maximized to $F(\theta=\pi/2)\approx 1$ where it is minimized. The higher $F(\theta)$ compared to the value for a radial photon trajectory, $\sin^2\theta$, is due to the generically higher transverse magnetic fields as well as longer propagation distances along non-radial photon trajectories.

The magnetic field of a white dwarf is usually detected by observing the associated Zeeman splittings and shifts of its spectral lines. For featureless (DC) white dwarfs such as blackbody stars, the magnetic fields can be detected through a different means, namely circular polarimetry. This method is sensitive to line-of-sight magnetic fields that are sufficiently strong ($\gtrsim 10^5\text{ G}$) to circularly polarize the photons at the level of $\gtrsim 10^{-4}$ \cite{2020A&A...643A.134B,2022A&A...657A.105B}. DC white dwarfs with magnetic fields $\sim 10^5-10^8\text{ G}$ have recently been discovered through circular polarimetry \cite{2020A&A...643A.134B,2022A&A...657A.105B}. Note that these are the average values of the magnetic fields along the line of sight; the magnetic fields at the pole $B_{\rm WD}$ can be significantly larger. These discoveries are in agreement with the statistical finding that about 20\% of white dwarfs posses strong magnetic fields. White dwarf magnetic fields observed thus far are distributed approximately uniformly in the log-space in the range $10^4-10^9\text{ G}$ \cite{2022A&A...657A.105B}. Based on these statistics, we can expect 3 out of the 17 white dwarfs to be magnetic in the aforementioned range.

\section{Data analysis and results}
\label{s:results}

\subsection{Data sets}

We use the spectrometric and photometric data of the 17 blackbody stars identified in \cite{2018AJ....156..219S}. For the spectrometric data, we use DR16 SkyServer \footnote{\url{https://skyserver.sdss.org/dr16}} to access the Catalog Archive Server (CAS) database of the SDSS. For each source, we use the \textit{science quality} data that are flagged as “primary” (see Table\,\ref{tbl:spec}, Appendix\,\ref{app:dataset} for details) and verify that in almost all cases these data have the smallest uncertainties and in some cases it also covers a wider range of measured wavelengths. These spectrometric data approximately cover wavelengths of $3500-10300\,\textup{\AA}$, with the best sensitivity at about $6000-10000\,\textup{\AA}$.

Since the SDSS optical observations cover only the longer wavelength region of the full spectrum of the stars of interest given their temperature, it is useful to incorporate UV photometric observations of the Galaxy Evolution Explorer (GALEX) telescope \cite{GALEX}, which in combination with the SDSS photometric data allows resolving the blackbody spectral peak and turnover. GALEX has pivot wavelengths (i.e. effective filter wavelengths) about $1536\,\textup{\AA}$ and $2299\,\textup{\AA}$ for its two FUV and NUV photometry bands. SDSS has five filtered photometry bands denoted as $u, g, r, i,$ and $z$ with the corresponding pivot wavelengths about $3557\,\textup{\AA}, 4702\,\textup{\AA}, 6176\,\textup{\AA}, 7490\,\textup{\AA},$ and $ 8947\,\textup{\AA}$, respectively. We obtain the photometry data from the SIMBAD astronomical database\footnote{\url{http://simbad.cds.unistra.fr/simbad/}} \cite{Wenger:2000sw}; see Table\,\ref{tbl:photometry}, Appendix\,\ref{app:dataset} for details.

\subsection{Quality of the blackbody spectra}
In the absence of BSM physics, we model the underlying spectrum emitted by a blackbody star with a perfect blackbody profile $B_\lambda(T)$. The apparent flux as seen in a telescope $F_{\lambda}$ is $\pi (R/d)^2$ times the emergent spectrum right outside the star. As the radius $R$ and distance $d$ to the star introduce new sources of errors, we fit away the normalization of the observed spectrum using a rescaled blackbody fit
\begin{align}
    aB_\lambda(T)=a\frac{4\pi}{\lambda^5}\frac{1}{e^{2\pi/\lambda T}-1}
\end{align}
 We determine the normalization $a$ and temperature $T$ by minimizing the chi-squared function constructed as follows
\begin{align}
    \chi^2_{B}=\sum_{i=1}^{N_{\rm bin}} \left(\frac{F_{i}-aB_{\lambda_i}(T)}{\sigma_{F_{i}}}\right)^2\label{chisqab}
\end{align}
where the sum is over all the $N_{\rm bin}$, which is the wavelength bins of the dataset in use, and $F_i$ and $\sigma_{F_i}$ are the observed flux in a wavelength bin $i$ and the associated error. The SDSS spectrometry datasets for a given star typically have $N_{\rm bin}\approx 4600$ which varies by $O(1)$ depending on the star and typical error bars of $\sigma_{F_i}/F_i\sim 10\%$, while the SDSS+GALEX photometry datasets have $N_{\rm bin}=5-7$ depending on whether the GALEX data are available for the star of interest and typical error bars of $\sigma_{F_i}/F_i\sim 1\%$.

All the 17 spectra of the blackbody stars fit well with a rescaled blackbody spectrum $aB_\lambda(T)$ with $(a,T)$ tuned to minimize $\chi^2_{B}$. The best-fit reduced chi-squared  $\left[\chi^2_{B}\right]_{\text{best}(a,T)}/N_{\rm dof}$ (where in this case $N_{\rm dof}=N_{\rm bin}-2$) for the spectrometry and photometry data lie, respectively, in the ranges $0.777-2.80$ and $0.409-15.1$,  with an average of $1.13$ and $5.82$. See Appendix~\ref{app:dataset} for more details about the fits. Oscillatory residuals $F_i-aB_{\lambda_i}(T)$ to the best-fit $aB_\lambda(T)$ function are seen but with no more than 2 nodes. While this seems like a characteristic signature of photon-axion conversions, residuals displaying such shapes are expected in the absence of BSM physics when fitting a convolved blackbody, reflecting a sum of contributions from different atmospheric layers, with a pure blackbody since the latter is always narrower than the former (see Appendix~\ref{appendix:gray}).

Given how well the base template $aB_\lambda(T)$ fits the data, we can rule out or even detect dark sector phenomena at varying levels of confidence based on how their inclusion affects the overall goodness of fit, as measured by the change in the chi-squared value. We discuss below how the blackbody stars can be used as axion probes.

\subsection{Axion limits}
In the presence of photon-axion conversion, the expected flux of the initially unpolarized light from a star is modified as
\begin{align}
    F_\lambda^{a}=\left[1-P_{\gamma\rightarrow a}(\lambda)\right]aB_\lambda(T)
\end{align}
where $P_{\gamma\rightarrow a}(\lambda)$ is photon to axion conversion probability computed in Section~\ref{s:signals}. Since the normalizations of the spectra are fitted away in our analysis, our technique is at present sensitive only to \textit{shape distortions}. As displayed in Figs.~\ref{fig:galconvprob} and \ref{fig:WDconvprob} the strongest conversion probabilities occur in the part of the parameter space where the spectral distortions vary monotonically with the wavelength. Despite their monotonicity, the detailed shapes of these spectra are non-trivial and are unlikely to be exactly mimicked by known astrophysical processes.

Our strategy in obtaining the axion exclusion limits is as follows. We construct the chi-squared function for the axion model $\chi^2_{a}$ in a way similar to the $\chi^2_{B}$ defined in \eqref{chisqab}
\begin{align}
    \chi^2_{a}=\sum_{i=1}^{N_{\rm bin}} \left(\frac{F_{i}-F^{a}_{\lambda_i}(m_a,g_{a\gamma\gamma},a,T)}{\sigma_{F_{i}}}\right)^2\label{chisqa}
\end{align}
For each axion mass $m_a$, we first minimize $\chi^2_{a}$ over the other three parameters $(g_{a\gamma\gamma}, a, T)$, giving the best-fit chi-squared of $\left[\chi^2_{a}\right]_{\text{best}(g_{a\gamma\gamma}, a,T)}$. Next, for the same $m_a$ we compute the best-fit chi-squared $\left[\chi^2_{a}(g_{a\gamma\gamma})\right]_{\text{best}(a,T)}$ for arbitrary $g_{a\gamma\gamma}$, now minimizing only over $(a,T)$. By Wilk's theorem \cite{Cowan:2010js}, we expect the difference between the aforementioned chi-squared values to follow a (one-sided) chi-squared distribution and thereby set the $95\%$ confidence limit on $g_{a\gamma\gamma}$ by the following criterion for a given $m_a$
\begin{align}
    \left[\chi^2_{a}(g_{a\gamma\gamma})\right]_{\text{best}(a,T)}-\left[\chi^2_{a}\right]_{\text{best}(g_{a\gamma\gamma},a,T)}>\chi^2_{95\%} \label{95CL}
\end{align}
where $\chi^2_{95\%}\approx 2.71$ for one degree of freedom.

\begin{figure*}
    \centering
    \includegraphics[width=0.75\linewidth]{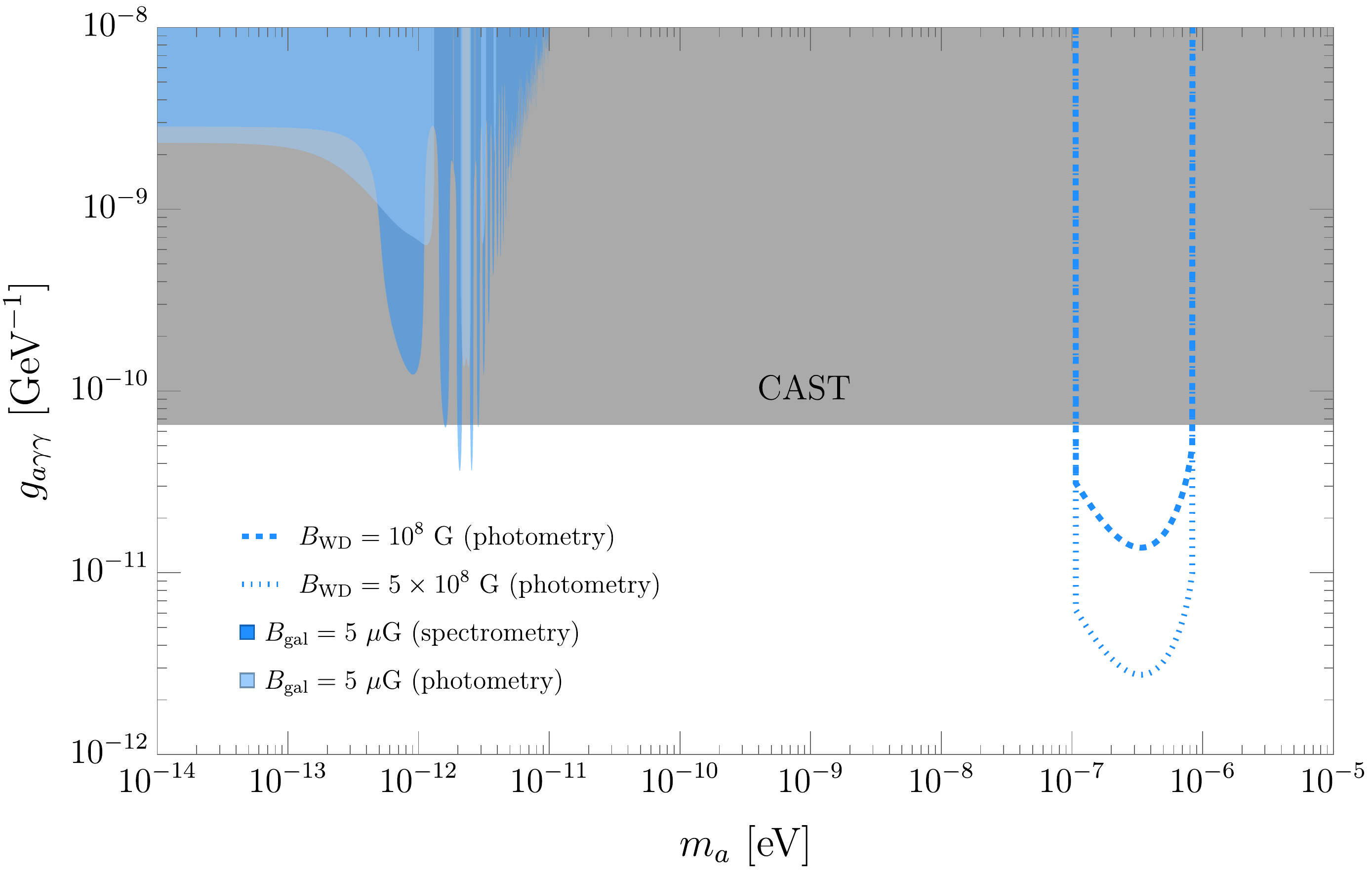}\\
    \caption{The 95\% confidence-level limits on the axion parameter space $(m_a,g_{a\gamma\gamma})$ obtained from both the spectrometric and photometric data of the blackbody star J004830.324+001752.80 which gives the best limits and whose properties can be found in Table.~\ref{tbl:bb_fit}. We derived these limits based on the lack of spectral distortions due to photon-axion conversions in the local galactic magnetic field (colored regions) and in putative stellar magnetic fields (dashed lines). While the magnetic field of this star has not been measured, we show the would-be exclusion limits for the fiducial cases where the star has polar magnetic fieldd $B_{\rm WD}=10^{8}\text{ G}$ and $B_{\rm WD}=5\times 10^{8}\text{ G}$, i.e. values consistent with existing magnetic field measurements of DC white dwarfs \cite{2020A&A...643A.134B}, and oriented such that $F(\theta)=2$ in \eqref{PWD}. For the limits based on photon-axion conversion in the galactic magnetic field $B_{\rm gal}$, we assume a constant local $B_{\rm gal}=5\,\mu\text{G}$ (c.f. \eqref{Bgal}) and a linearly-varying plasma-mass-squared profile $\omega_p^2$ along the line of sight (c.f. \eqref{plasmamassprofile}) with $L_p=+200\text{ pc}$.}

    \label{fig:parameterspace}
\end{figure*}

Fig.~\ref{fig:parameterspace} shows the limits derived from the spectrometry and photometry data of the blackbody star that yields the most stringent limits.\footnote{The limits from the other blackbody stars are shown in Fig.~\ref{fig:parameterspaceapp}.} We found that, in accordance with our expectations, the best limits occur at around $m_a\sim 10^{-12}\text{ eV}$ and $m_a\sim \mu\text{eV}$ for photon-axion conversions via the galactic and stellar magnetic fields, respectively. The limits that rely on conversions in the galactic magnetic field are currently inferior to other limits in the same mass range (see e.g. \cite{Payez:2014xsa,Berg:2016ese,Marsh:2017yvc, Chen:2017mjf,Reynolds:2019uqt,Xiao:2020pra,Payez_2012}), nevertheless they can improve significantly in the future. On the other hand, the projected limits based on anticipated future measurements of the star's magnetic field with the fiducial values at the pole $B_{\rm WD}=10^{8}\text{ G}$ and $B_{\rm WD}=5\times 10^{8}\text{ G}$ can reach $g_{a\gamma\gamma}\lesssim 10^{-11}\text{ GeV}^{-1}$ and $g_{a\gamma\gamma}\lesssim 3\times 10^{-12} \text{ GeV}^{-1}$ respectively, which  are competitive with the current best limits in the same mass range \cite{Dessert:2022yqq,Noordhuis:2022ljw}.\footnote{These limits, including ours, are of course subject to various systematics. New arrival of data as well as progress in our understanding of the pertaining astrophysical systems will determine the future potentials of the proposed techniques.}  We stress that the limits on $g_{a\gamma\gamma}$ are highly dependent on the assumed values of $B_{\rm WD}$. The $B_{\rm WD}$ values we adopt here are consistent with typical circular-polarimetry measurements of DC white dwarf magnetic fields along the line of sight \cite{2020A&A...643A.134B}, but it can in principle be much weaker or much stronger.\footnote{To account for higher $B_{\rm WD}$ than the assumed values, we would need to take into account strong QED effects as done in e.g. \cite{Chelouche:2008ta, Dessert:2022yqq}, which we leave for future works. For $B_{\rm WD}\leq 5\times 10^8\text{ G}$ assumed in our analysis, strong QED effects result in negligible ($\leq 10\%$) changes in the photon-axion conversion probability for axion masses $m_a\geq 8\times 10^{-8}\text{ eV}$.}

In obtaining the aforementioned limits, we have avoided axion mass ranges where the boundary of the excluded regime \eqref{95CL} yields $P_{\gamma\rightarrow a}\geq 10\%$ anywhere in the relevant wavelength window, as the approximate probability formula \eqref{weakmixing} becomes inaccurate in that regime. This results in $m_a$ gaps in our $B_{\rm gal}$-based limits and set the $m_a$ range of our $B_{\rm WD}$-based limits. In the case of $B_{\rm gal}$-induced photon-axion conversion, we found that for certain axion masses both the spectrometry and photometry data exhibit significant preference toward the existence of axion with $g_{a\gamma\gamma}\neq 0$. Nevertheless, given the simplicity of our base spectrum $aB_\lambda(T)$ this is most likely due to our mismodelling of the blackbody-star spectra. Hence, rather than searching for imprints of the axion we focus on placing limits on them by  replacing $\left[\chi^2_{a}\right]_{\text{best}(g_{a\gamma\gamma},a,T)}$ in \eqref{95CL} with $\left[\chi^2_{a}(g_{a\gamma\gamma}=0)\right]_{\text{best}(a,T)}$ which is equivalent to $\left[\chi^2_B\right]_{\text{best}(a,T)}$. Incidentally, the SDSS spectrometer, due to its limited wavelength range, probes only the parts of the blackbody-star spectra on the longer wavelength side of the blackbody peaks that are entirely monotonic. Since the $B_{\rm WD}$-induced photon-axion conversion probability is monotonic with respect to the wavelength, this leads to a degeneracy between the rescaled blackbody spectrum parameters $(a,T)$ and the photon-axion conversion effect. The UV data points, located on the shorter wavelength side of the blackbody peaks, from GALEX are essential to break this degeneracy, and for this reason we show only $B_{\rm WD}$-based limits that are obtained with the photometric data in Fig.~\ref{fig:parameterspace}.

\subsection{Future prospects}
\label{ss:future}

We first discuss how the limits derived from the blackbody stars considered here might improve in the future with better measurements. All the fluxes reported here are what remain after the sky background has been subtracted. Since the distances of the blackbody stars are $\lesssim 100\text{ pc}$, the reddening due to interstellar-medium extinction is negligible \cite{2018AJ....156..219S}. Other possibly important sources of error in the observed photon fluxes include intrinsic instrumental noise and Poisson counting noise\footnote{The Poisson noise appears to be dominated by the subtracted sky background flux, $F_{\lambda}^{\rm sky}\sim 10^{-17}-10^{-15}\text{erg/s}/\text{cm}^2/\textup{\AA}$, which is about $O(1-10)$ larger than the signal flux. Assuming $2.5\text{ m}$ telescope diameter and 15 minutes exposure time, the number of photons arriving in each $\Delta \lambda\sim 1.5\textup{\AA}$-wide wavelength bin can be estimated as $N_{\gamma}=F_\lambda\Delta \lambda A_{\rm det} t_{\rm det}$. The expected fractional error in intensity due to Poisson fluctuations is found to be $\sqrt{N_{\gamma}^{\rm sky}}/N_\gamma^{\rm signal}\sim 0.05-0.5$, i.e. $\sim 10\%$ error in the observed intensity, consistent with the reported error bars.}. Both these noises would improve with follow-up observations using better and dedicated telescopes. Current and planned optical telescopes with better exposures and/or resolutions include Gemini \cite{2022ApJ...934...36B}, HIRES at Keck \cite{1994SPIE.2198..362V}, and NIRSpec at JWST \cite{2020ApJ...901L...1K}. Future missions can also extend the wavelength window of the spectroscopic analysis to outside the optical range and, moreover, improve the distance and stellar radius inferences perhaps to a level where we can use the normalizations of the spectra as an extra handle.

Once various measurement-related noises are suppressed, the remaining limiting factor is the quality of the blackbody itself. As per our current understanding, the blackbody stars are white dwarfs with effective temperatures in the range $6000-11000\text{ K}$, composed mainly of helium, and possibly contains a trace amount of hydrogen. The physics potential of our proposed technique will depend on the likelihood of detecting such stars with future surveys.

An advantage of using DC white dwarfs as a backlight is that they are by definition uncluttered by strong absorption lines. However, this also means that their atmospheric properties must be inferred solely from matching the predictions of a model atmosphere with the observed continuum spectrum. On the other hand, the presence of spectral lines can provide additional information that help pin down the properties of the atmosphere, thus suggesting that there is the potential that white dwarfs of other spectral types, e.g. DB, might serve as an even better backlight than DC ones if, aside from the absorption lines, part of their spectra show a blackbody-like profile.

On the modelling side, improvements can also be made with new data. Better modelling of the photon-axion conversion requires better a knowledge of the magnetic field and plasma mass in the vicinity of the Earth and the blackbody stars. The wavelength-dependent photon deficit due to conversion to axion mimics the effect of continuum opacity in the white dwarf atmosphere. Both these effects vary from one white dwarf to another in a way that is difficult to pin down if the white dwarfs are not fully characterised. However, one can in principle break this partial degeneracy between the axion and opacity effects based on their different wavelength dependencies. This would require flux measurements with sufficiently high signal to noise ratio as well as a careful modelling of the white dwarf atmosphere, which we leave for future work. One easily-implementable step toward a more realistic model-atmosphere is by adopting as the base fit the (rescaled) spectrum predicted by the gray atmosphere model (discussed in Appendix~\ref{appendix:gray}, which accounts for atmospheric stratification effects, instead of the pure blackbody spectrum used in the main text).

\section{Discussion and conclusion}
\label{s:conclusion}
We proposed using \textit{blackbody stars}, 17 stars in the SDSS catalog with properties consistent with a helium-rich DC white dwarfs and exhibiting near-perfect blackbody spectra in the optical range, as probes of BSM physics. We consider as a reference the model of an axion with coupling to photons, which predicts photon-axion conversions with wavelength-dependent probabilities in the presence of (galactic or stellar) background magnetic fields, as illustrated in Figs.~\ref{fig:galconvprob} and \ref{fig:WDconvprob}. Based on how the inclusion of these effects worsens the overall goodness-of-fit to the observed blackbody star spectra, we derive competitive projected limits on the axion parameter space, displayed in Fig.~\ref{fig:parameterspace}, demonstrating the feasibility of this approach. Our proposed technique will benefit from both the addition of new blackbody stars and the improved precision of the spectra data of these stars.

The spectrum of photons emerging from a blackbody star is unprotected against spectral distortions occurring between the decoupling of photons at the photosphere and their arrival at the telescope. We have focused on distortions affecting these decoupled photons. The ``placement" of this effect is analogous to the Sunyaev-Zeldovich effects on the decoupled CMB. Since a BSM mechanism must affect the decoupled optical photons directly in order to cause distortions in the optical window, the range of BSM models that one can probe is somewhat limited. To extend the analogy with how the CMB is used as a BSM probe, it would be interesting to explore the prospects of constraining a wide range of BSM effects operating at energies outside the optical range that does not entail complete rethermalization and instead get reprocessed into spectral distortions in the optical range. This may occur if the BSM effects take place below the photosphere (where the atmosphere is more opaque) in a variety of ways, such as electromagnetic injections affecting the radiative transfer equation, alterations of the atmospheric excitation and ionization equilibrium, and anomalous opacity effects e.g. due to the addition of new contaminants.

\section*{Acknowledgments}
We are very grateful to Gautham A. P., Christopher Dessert, Peter Graham, David E. Kaplan, Surjeet Rajendran, Kevin Schlaufman, and Nao Suzuki for fruitful exchanges. JHC is supported in part by the NSF grant PHY-2210361, the Maryland Center for Fundamental Physics, and JHU Joint Postdoc Fund. RE is supported in part by the University of Maryland Quantum Technology Center.

This research made use of \texttt{Mathematica} \cite{Mathematica}, \texttt{Jupyter} \cite{soton403913}, \texttt{NumPy} \cite{2020Natur.585..357H}, \texttt{SciPy} \cite{2020NatMe..17..261V}, \texttt{matplotlib} \cite{2007CSE.....9...90H}, \texttt{Pyphot} \cite{Fouesneau_pyphot_2022}, \texttt{Pandas} \cite{pandas}, and \texttt{Astropy} \cite{astropy:2022}. This research has made use of the SIMBAD database,
operated at CDS, Strasbourg, France \cite{Wenger:2000sw}.

\appendix

\section{Gray atmosphere model}
\label{appendix:gray}

The flux of starlight emerging into the empty space is the sum of contributions from many stellar atmospheric layers with varying temperatures, pressures, and compositions. Hence, one might expect deviations from a single-temperature blackbody spectrum even in the absence of frequency-dependent opacity effects discussed in the main text. We argue here that in cases of our interest this type of deviations are small. 

Most of the contributions to the emergent flux would come from by the layers near the photosphere, where the opacity is order unity. The radiation intensities coming from the outer layers is weaker due to the lower temperatures, while those coming from the deeper layers are strongly suppressed by opacity effects, despite their higher temperatures. The resulting spectrum can resemble a pure blackbody if the variation in the properties of the dominant emitting layers around the photosphere is minimal. That said, some broadening of the spectrum is in general expected. To quantify this broadening effect, we consider a toy atmosphere model with plane-parallel geometry and opacity assumed frequency-independent, i.e. the so-called gray atmosphere model \cite{1963ApJ...138..281O, 1978stat.book.....M,2014tsa..book.....H}. The radiative transfer equation in this case is considerably more tractable and commonly discussed in many textbooks. We highlight below several key results relevant to our discussion.

The temperature as a function of optical depth $\tau$ in the gray atmosphere model is given by \cite{1978stat.book.....M,2014tsa..book.....H}
\begin{align}
    T_{\rm gray}^4(\tau)=\frac{3}{4}\left[\tau+q(\tau)\right]T_{\rm eff}^{4} \label{Tgray}
\end{align}
The optical depth at an atmospheric layer with a vertical location $z$ (the precise reference point of $z$ is unimportant) is defined in terms of the atmospheric opacity $\kappa$ (the reciprocal photon-absorption mean free path) as $\tau(z)=\int_z^\infty \kappa dz'$. The effective temperature $T_{\rm eff}$ is the temperature of a pure blackbody that yields a given luminosity $L$ of a star, defined by the relation $\sigma T_{\rm eff}^4=L/4\pi R^2$ where $\sigma=\pi^2/60$ is the Stefan-Boltzmann constant. The so-called Hopf function $q(\tau)$ has no known closed analytical form, but is well-approximated to percent level by the following function
\begin{align}
    q(\tau)\approx 1-\frac{\sqrt{3}}{6}+\left(\frac{\sqrt{3}}{2}-1\right)e^{-2\sqrt{3}\tau}
\end{align}
The emergent monochromatic photon flux at $\tau=0$ is a weighted sum of the blackbody emission from different layers\footnote{This is the flux of photon energy through a unit surface right outside the star. The flux arriving at a telescope is $\pi (R/d)^2$ times this flux.} \cite{1978stat.book.....M,2014tsa..book.....H}
\begin{align}
    \left(\frac{dL}{d\lambda}\right)_{\rm gray}=2\int_0^\infty d\tau B_\lambda\left[T_{\rm gray}(\tau)\right]E_2(\tau) \label{grayflux}
\end{align}
where $E_n(x)$ is the exponential integral, defined as
$E_n(x)\equiv x^{n-1}\int_x^\infty dy\;e^{-y}/y^n$.

\begin{figure}
    \centering
\includegraphics[width=0.915\linewidth]{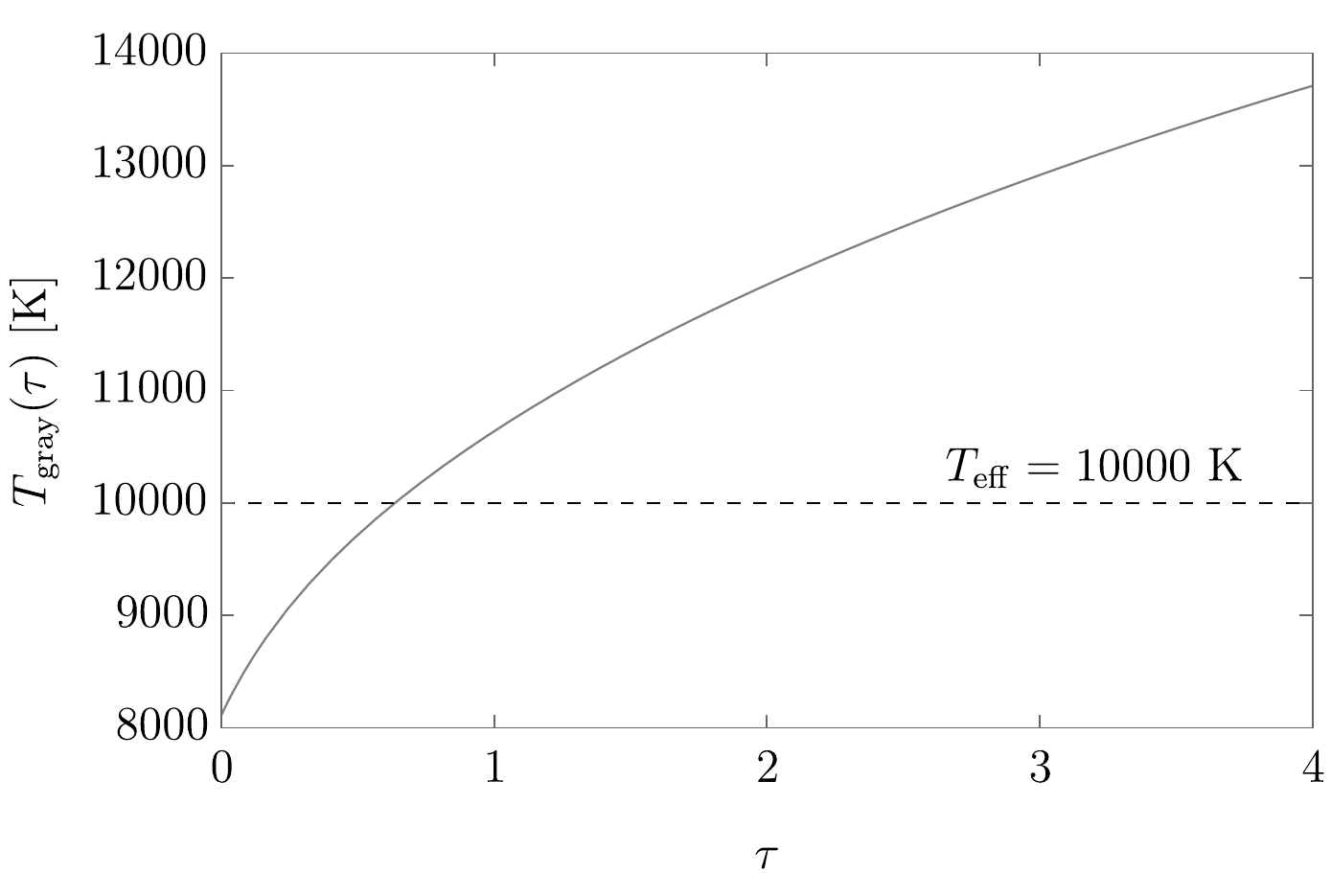}\\
    \includegraphics[width=\linewidth]{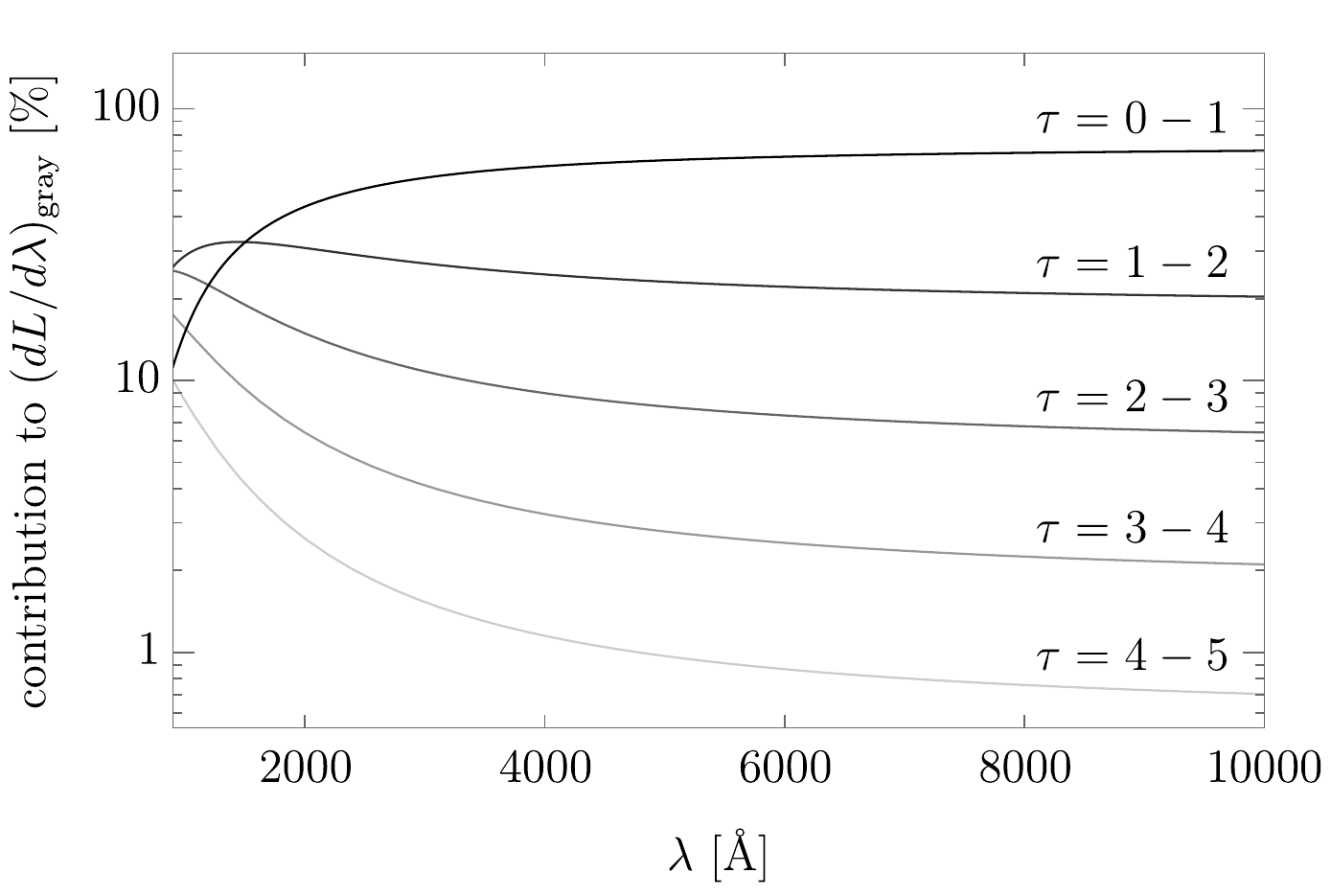} 
   \caption{\textit{Top: }temperature $T_{\rm gray}(\tau)$ as a function of optical depth $\tau$ predicted by the gray atmosphere model, c.f.~\eqref{Tgray}, for $T_{\rm eff}=10^4\text{ K}$. \textit{Bottom: }the fractional contributions to the emergent flux in the gray atmosphere model $(dL/d\lambda)_{\rm gray}$ from different ranges of the optical depth. About $99\%$ of the contributions to $(dL/d\lambda)_{\rm gray}$ in the wavelength range $\lambda=3500-10300\textup{\AA}$ come from optical depths $\tau<4$.} 
    \label{fig:graytemp} 
\end{figure}

\begin{figure}
    \centering
    \includegraphics[width=\linewidth]{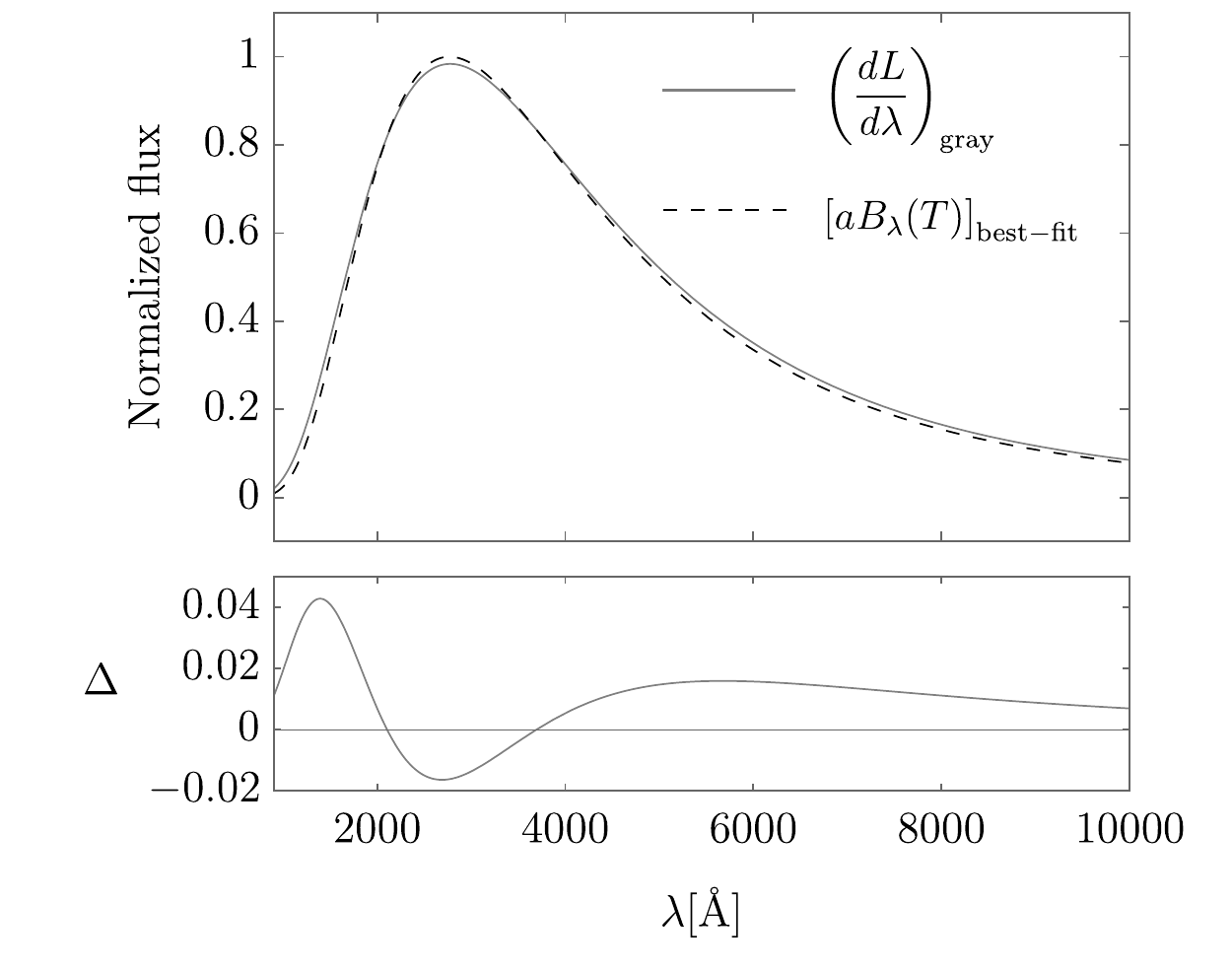}
    \caption{\textit{Top: }the emergent flux in the gray atmosphere model $(dL/d\lambda)_{\rm gray}$, c.f.~\eqref{grayflux}, for $T_{\rm eff}=10^4\text{ K}$ shown together with rescaled blackbody spectrum $aB_{\lambda}(T)$ that fits $(dL/d\lambda)_{\rm gray}$ best over the wavelength range $1000-10000\textup{\AA}$. \textit{Bottom: }the difference between the spectrum predicted in the gray atmosphere model and the best-fit rescaled pure blackbody spectrum $\Delta \equiv (dL/d\lambda)_{\rm gray}-aB_{\lambda}(T)$ (normalized relative to the peak).}
    \label{fig:grayspec}
\end{figure}

Fig.~\ref{fig:graytemp} shows the temperature as a function of optical depth predicted by the gray atmosphere model as well as the relative contributions to the emergent spectrum from different layers. The dominant contributors in the wavelength range $3500-10300\textup{\AA}$ come from the optical depths $\tau\lesssim 2$, over which the temperature variation is $\lesssim 20\%$. For this reason, the spectrum emerging from a gray atmosphere is well-fit by a rescaled pure blackbody spectrum. Fig.~\ref{fig:grayspec} shows that the best-fit rescaled blackbody spectrum $aB_\lambda(T)$ agrees with $(dL/d\lambda)_{\rm gray}$ at few-percent level. Since the sum of blackbody emissions from different layers is always broader than a pure blackbody with a similar average energy, the residual to the $aB_\lambda(T)$ fit may show oscillatory features with one or two nodes. We found that using the spectrum  predicted in the gray-atmosphere model (in the absence of axion) to fit the spectrometry data give chi-squared values that deviate from the pure-blackbody ones by $\lesssim 0.1$.

\section{Orientation dependence of photon-axion conversion in the stellar magnetic field}
\label{appendix:nonradial}
\begin{figure}
    \centering
    \includegraphics[width=\linewidth]{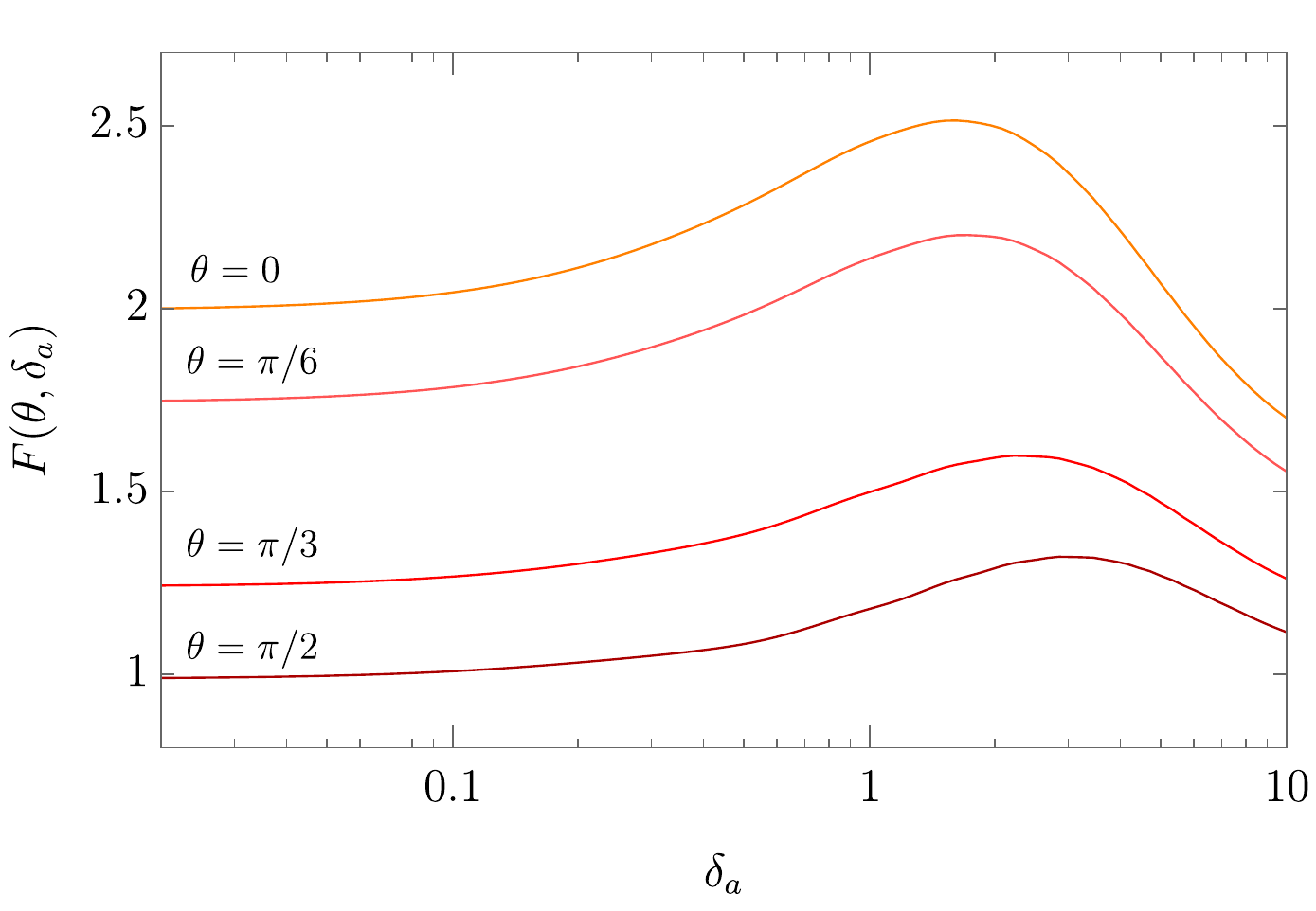}
    \caption{The numerical factor $F(\theta)$ as defined in \eqref{PWD}, which in principle depends also on $\delta_a$. The results show that $F(\theta)$ varies from $\approx 1$ to $\approx 2$ as $\theta$ is varied from $0$ to $\pi/2$ and depends mildly on $\delta_a$.}
    \label{fig:Ftheta}
\end{figure}

We computed the photon-axion conversion probability in the stellar magnetic field for non-radial trajectories followed by photons emitted from different parts of the stellar surface facing the telescope, considering both linear polarizations of the photons. We then average these probabilities over the cross-section of the star facing the telescope. The results are shown in Fig.~\ref{fig:Ftheta}. This justifies the assumption we made in the main analysis that $F(\theta)$ depends mildly on $\delta_a$ and varies in the range $F(\theta)\approx 1-2$ for $\theta=0-\pi/2$.

\section{Blackbody fitting of the photometric data}\label{app:photmetry_fitting}
In the fitting routine of the photometry data, we use \textit{filtered} rescaled blackbody and ``rescaled blackbody + axion'' spectrum $ F_\lambda^{a}=\left[1-P_{\gamma\rightarrow a}(\lambda)\right]aB_\lambda(T)$ as our model prediction. We utilize the \texttt{Pyphot} package \cite{Fouesneau_pyphot_2022} to apply GALEX as well as SDSS photometry filters to the model spectrum which results in seven flux functions which depend on the model parameters at each pivot wavelength.

To perform this analysis numerically, we compute the filtered spectra for a range of temperatures and axion masses. The resultant photometry fluxes at pivot wavelengths are then interpolated to achieve a prediction for the photometry observations. In order to reduce computational costs we first approximate  the predicted spectra as
\begin{align}
    aB_\lambda(T)\left[1-P_{\gamma\rightarrow a}(\lambda)\right] \simeq & aB_\lambda(T) - aB_\lambda(T_\ast) P_{\gamma\rightarrow a}(\lambda) \nonumber\\
    &- a(T-T_\ast)\dfrac{\partial B_\lambda}{\partial T}\bigg|_{T=T_\ast} P_{\gamma\rightarrow a}(\lambda)
\end{align}
where we essentially Taylor expand the blackbody spectrum around the best-fit temperature $T_\ast$ (in the absence of axion). In this case filtered form of the first (second) term on the right hand side only needs to be interpolated as a function of temperature (axion mass). We numerically verify that the interpolated function using the above method agrees with the filtered flux computed using the full form of the signal flux.

\section{Dataset}\label{app:dataset}
We provide the spectrometric and photometric data used in our analysis in Table\,\ref{tbl:spec} and Table\,\ref{tbl:photometry}, respectively. To convert the AB magnitude data of Table\,\ref{tbl:photometry} to flux data we use the following relation:
\begin{equation}
    \mathrm{flux} = \dfrac{c}{\lambda^2_\mathrm{p}}10^{-(\mathrm{AB~ mag.}+48.6)/2.5}
\end{equation}
where $\lambda_\mathrm{p}$ is the pivot wavelength of the corresponding bandpass filter; $c$ is the speed of light; and the flux is given in the units of $\mathrm{erg\,s^{-1}\,cm^{-2}\,\textup{\AA}^{-1}}$. We also provide the best fit values of blackbody fits using both spectrometric and photometric data in Table\,\ref{tbl:bb_fit}. Distances and radii in Table\,\ref{tbl:bb_fit} are adapted from ref.\,\cite{2019A&A...623A.177S}.

\begin{table*}[h!]
\begin{center}
\begin{tabular}{l|c|c||c|c|c|c}
SDSS name& ~~~RA (J2000)~~~ & ~~~DEC (J2000)~~~& ~~~Plate~~~ & ~~~MJD~~~ & ~~~Fiber~~~ & ~~~Flag~~~\\\hline
J002739.497-001741.93&00:27:39.497   &-00:17:41.93   &  4220 & 55447 & 0270 & Primary\\
J004830.324+001752.80&00:48:30.324   &+00:17:52.80   &  3590 & 55201 & 0726 & Primary\\
J014618.898-005150.51&01:46:18.898   &-00:51:50.51   &  4231 & 55444 & 0054 & Primary\\
J022936.715-004113.63&02:29:36.715   &-00:41:13.63   &  4238 & 55455 & 0226 & Primary\\
J083226.568+370955.48&08:32:26.568   &+37:09:55.48   &  3762 & 55507 & 0908 & Primary\\
J083736.557+542758.64&08:37:36.557   &+54:27:58.64   &  5156 & 55925 & 0850 & Primary\\
J100449.541+121559.65&10:04:49.541   &+12:15:59.65   &  5328 & 55982 & 0184 & Primary\\
J104523.866+015721.96&10:45:23.866   &+01:57:21.96   &  4733 & 55649 & 0346 & Primary\\
J111720.801+405954.67&11:17:20.801   &+40:59:54.67   &  4651 & 56008 & 0608 & Primary\\
J114722.608+171325.21&11:47:22.608   &+17:13:25.21   &  5892 & 56035 & 0888 & Primary\\
J124535.626+423824.58&12:45:35.626   &+42:38:24.58   &  4702 & 55618 & 0958 & Primary\\
J125507.082+192459.00&12:55:07.082   &+19:24:59.00   &  5859 & 56065 & 0664 & Primary\\
J134305.302+270623.98&13:43:05.302   &+27:06:23.98   &  6002 & 56104 & 0747 & Primary\\
J141724.329+494127.85&14:17:24.329   &+49:41:27.85   &  6746 & 56386 & 0984 & Primary\\
J151859.717+002839.58&15:18:59.717   &+00:28:39.58   &  4012 & 55327 & 0514 & Primary\\
J161704.078+181311.96&16:17:04.078   &+18:13:11.96   &  4073 & 55663 & 0580 & Primary\\
J230240.032-003021.60&23:02:40.032   &-00:30:21.60   &  4207 & 55475 & 0102 & Primary
\end{tabular}
\caption{Plate, MJD, and Fiber labels of the SDSS spectrometric data used in this study. \label{tbl:spec}}
\end{center}
\end{table*}

\begin{table*}[h!]
\begin{center}
\begin{tabular}{l|c|c|c|c|c|c|c}
SDSS name & GALEX FUV& GALEX NUV& SDSS $u$& SDSS $g$& SDSS $r$& SDSS $i$& SDSS $z$\\\hline
J002739.497-001741.93  & 21.80$\pm$0.06 & 19.89$\pm$0.02 &  19.035$\pm$0.020 &	18.876$\pm$0.009 &	18.959$\pm$0.011 &	19.118 $\pm$0.014 & 19.340$\pm$0.048\\
J004830.324+001752.80  & 21.18$\pm$0.05 & 19.14$\pm$0.01 &  18.287$\pm$0.015 &	18.119$\pm$0.006 &	18.231$\pm$0.008 &	18.375$\pm$0.010 &	18.639$\pm$0.033\\
J014618.898-005150.51  & 20.61$\pm$0.05 & 18.80$\pm$0.01 &  18.219$\pm$0.014 &	18.117$\pm$0.007 &	18.241$\pm$0.008 &	18.420$\pm$0.010 &	18.616$\pm$0.030\\
J022936.715-004113.63  & 23.04$\pm$0.14 & 20.73$\pm$0.01 &  19.372$\pm$0.024 &	19.050$\pm$0.010 &	19.085$\pm$0.011 &	19.175$\pm$0.014 &	19.333$\pm$0.046\\
J083226.568+370955.48  &   &   & 19.390$\pm$0.028 &	18.971$\pm$0.10 &	18.853$\pm$0.011 &	18.834$\pm$0.013 &	18.928$\pm$0.039\\
J083736.557+542758.64  &   & 21.18$\pm$0.07 & 19.225$\pm$0.024 &	18.711$\pm$0.008 &	18.532$\pm$0.009 &	18.519$\pm$0.012 &	18.493$\pm$0.032\\
J100449.541+121559.65  &   &   & 19.393$\pm$0.024 &	19.168$\pm$0.010 &	19.210$\pm$0.012 &	19.305$\pm$0.015 &	19.413$\pm$0.056\\
J104523.866+015721.96  &   & 20.51$\pm$0.09 & 19.317$\pm$0.028 &	19.098$\pm$0.010 &	19.015$\pm$0.012 &	19.077$\pm$0.016 &	19.280$\pm$0.072\\
J111720.801+405954.67  & 20.93$\pm$0.34 & 18.98$\pm$0.09 &  18.246$\pm$0.015 &	18.060$\pm$0.006 &	18.173$\pm$0.007 &	18.322$\pm$0.009 &	18.617$\pm$0.032\\
J114722.608+171325.21  &   & 19.91$\pm$0.06 & 18.896$\pm$0.021 &	18.656$\pm$0.008 &	18.660$\pm$0.009 &	18.844$\pm$0.013 &	19.000$\pm$0.044\\
J124535.626+423824.58  &   & 18.30$\pm$0.03 & 17.304$\pm$0.009 &	17.110$\pm$0.004 &	17.181$\pm$0.005 &	17.283$\pm$0.006 &	17.445$\pm$0.015\\
J125507.082+192459.00  &   &  19.93$\pm$0.08 & 18.823$\pm$0.037 &	18.521$\pm$0.018 &	18.450$\pm$0.015 &	18.500$\pm$0.015 &	18.619$\pm$0.033\\
J134305.302+270623.98 &   &  20.03$\pm$0.14 & 19.054$\pm$0.022 &	18.922$\pm$0.008 &	19.008$\pm$0.014 &	19.142$\pm$0.016 &	19.339$\pm$0.065\\
J141724.329+494127.85  &  20.73$\pm$0.29 & 18.27$\pm$0.06 & 17.371$\pm$0.009 &	17.213$\pm$0.005 &	17.298$\pm$0.005 &	17.428$\pm$0.006 &	17.605$\pm$0.015\\
J151859.717+002839.58  &   &   &  19.746$\pm$0.028 &	19.430$\pm$0.011 &	19.368$\pm$0.013 &	19.495$\pm$0.019 &	19.573$\pm$0.057\\
J161704.078+181311.96  &   & 20.60$\pm$0.12 & 19.177$\pm$0.025 &	18.788$\pm$0.008 &	18.733$\pm$0.009 &	18.806$\pm$0.012 &	18.833$\pm$0.039\\
J230240.032-003021.60  &  20.79$\pm$0.06 & 18.88$\pm$0.01 & 17.968$\pm$0.013 &	17.794$\pm$0.006 &	17.893$\pm$0.007 &	18.033$\pm$0.008 &	18.254$\pm$0.22 
\end{tabular}
\caption{Photometric data for the blackbody stars. The pivot wavelengths for FUV, NUV, $u, g, r, i,$ and $z$ bandpass filters are as follows: $1536.405\,\textup{\AA}$ and $2299.245\,\textup{\AA}, 3556.524\,\textup{\AA}, 4702.495\,\textup{\AA}, 6175.579\,\textup{\AA}, 7489.977\,\textup{\AA},$ and $ 8946.710\,\textup{\AA}$. \label{tbl:photometry}}
\end{center}
\end{table*}

\begin{table*}[h!]
\begin{center}
\begin{tabular}{l|c|c|c|c|c}
SDSS name & 
$T$ [K] & $a[\times10^{-23}]$ & $\chi^{2}$/dof & $D$ [pc] & $R$ [km]\\
\hline
J002739.497-001741.93 &
$10623\pm49$ & $0.448\pm0.007$& 1.8 & $229\pm19$ & $9705\pm780$ \\
J004830.324+001752.80 & 
$10638\pm30$ & $0.887\pm0.009$& 13.0 & $137\pm4$ & $8224\pm250$\\
J014618.898-005150.51 & 
$11749\pm37$ & $0.679\pm0.007$& 13.0 & $180\pm9$ & $ 9547\pm471$\\
J022936.715-004113.63  &
$8874\pm24$ & $0.660\pm0.008$& 15.1 & $155\pm8$ & $8332\pm480$\\
J083226.568+370955.48  &
$8037\pm72$ & $1.092\pm0.034$& 3.3 & $118\pm5$ & $8721\pm411$\\
J083736.557+542758.64  &
$7492\pm45$ & $1.821\pm0.041$& 7.2 & $91\pm2$ & $8627\pm298$\\
J100449.541+121559.65  &
$9877\pm112$ & $0.434\pm0.015$& 1.4 & $222\pm23$ &  $9630\pm995$\\
J104523.866+015721.96 & $8958\pm95$ & $0.663\pm0.023$ & 3.4& & \\
J111720.801+405954.67  &
$10955\pm79$ & $0.861\pm0.017$& 8.5 & $134\pm5$ & $7906\pm330$\\
J114722.608+171325.21  &
$9924\pm80$ & $0.686\pm0.017$& 5.5 & $157\pm10$ & $8342\pm531$\\
J124535.626+423824.58  &
$10192\pm40$ & $2.601\pm0.030$& 7.8 & $71\pm1$ & $7475\pm97$\\
J125507.082+192459.00  &
$8869\pm105$ & $1.165\pm0.042$ & 1.7 & $118\pm3$ & $8526\pm256$\\
J134305.302+270623.98  &
$10722\pm126$ & $0.422\pm0.015$& 0.4 & $176\pm11$ & $7387\pm460$\\
J141724.329+494127.85  &
$10564\pm48$ & $2.114\pm0.027$& 4.1 & $88\pm1$ & $8295\pm138$\\
J151859.717+002839.58  &
$9027\pm106$ & $0.465\pm0.018$& 1.6 & $170\pm14$ & $7662\pm622$\\
J161704.078+181311.96  &
$8721\pm69$ & $0.940\pm0.024$& 3.7 & $112\pm2$ & $7561\pm283$\\
J230240.032-003021.60  &
$10493\pm28$ & $1.245\pm0.011$& 7.6 & $122\pm2$ & $8710\pm196$
\end{tabular}
\caption{Blackbody fit results using photometry data. The distances and radii are adapted from ref.\,\cite{2019A&A...623A.177S}.}\label{tbl:bb_fit}
\end{center}
\end{table*}

\begin{figure*}[h]
    \centering
    \includegraphics[width=0.75\linewidth]{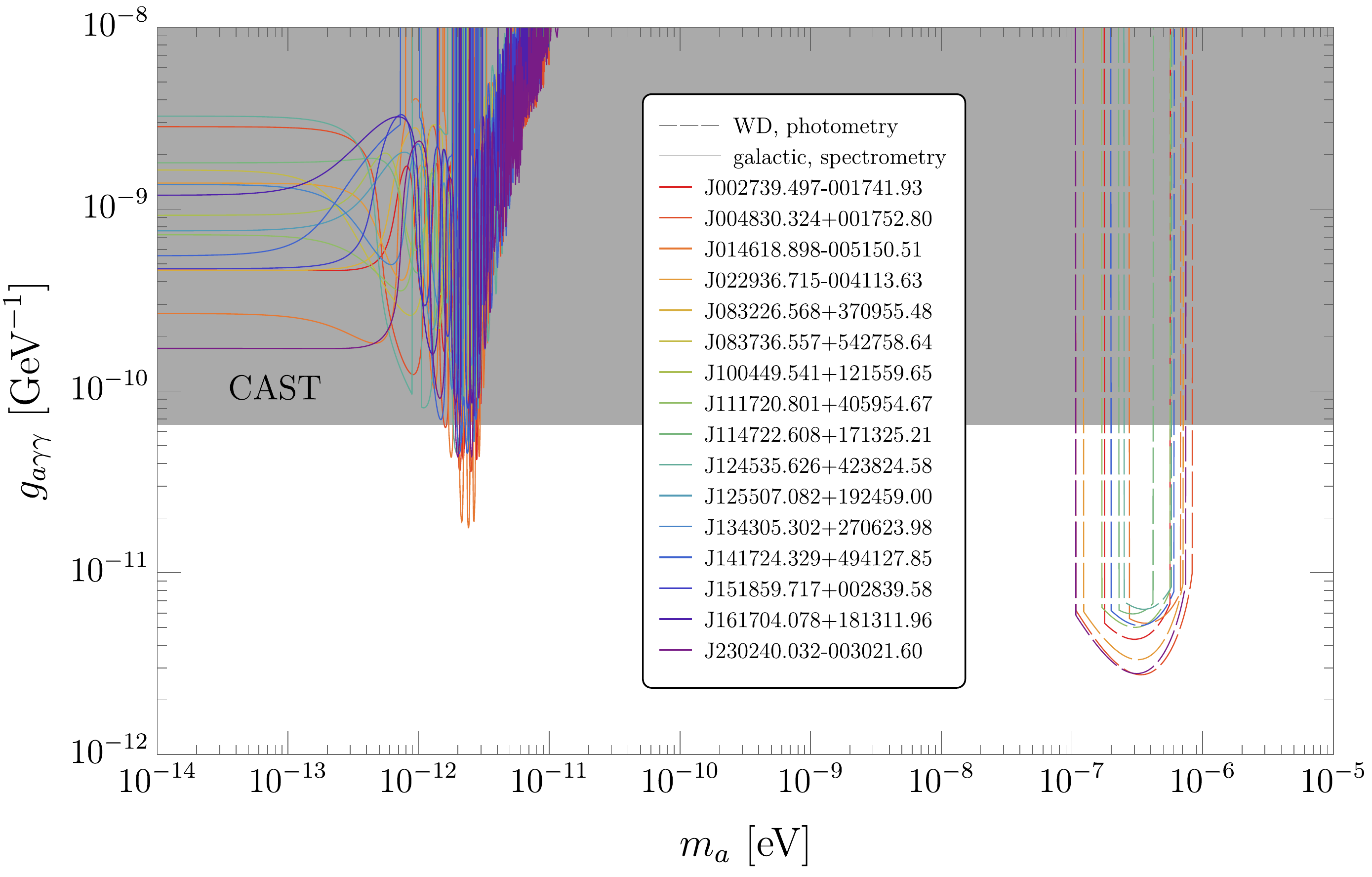}\\
\caption{The 95\% confidence-level limits on the axion parameter space $(m_a,g_{a\gamma\gamma})$ from the 16 (out of 17) blackbody stars with known distances and radii \cite{2019A&A...623A.177S}. The properties of these stars are collected in Table.~\ref{tbl:bb_fit}. The dashed and solid lines show, respectively, the limits that rely on photon-axion conversions in the stellar and galactic magnetic fields. While the magnetic fields of the blackbody stars have not been measured, we show the would-be exclusion limits if these stars possess a polar magnetic field $B_{\rm WD}=5\times 10^{8}\text{ G}$, a value consistent with existing magnetic field measurements of DC white dwarfs \cite{2020A&A...643A.134B}, and assume that the orientation of the star is such that $F(\theta)=2$ in \eqref{PWD}. For the limits from galactic photon-axion conversions, we assume a constant local $B_{\rm gal}=5\,\mu\text{G}$ (c.f. \eqref{Bgal}) and a linearly-varying plasma-mass-squared profile $\omega_p^2$ along the line of sight (c.f. \eqref{plasmamassprofile}) with $L_p=+200\text{ pc}$.}    \label{fig:parameterspaceapp}

\end{figure*}

\bibliography{references}
\bibliographystyle{apsrev4-1}

\end{document}